# Zinc Uptake and Radial Transport in Roots of *Arabidopsis thaliana:* A Modelling Approach to Understand Accumulation

Juliane Claus[1,2], Ansgar Bohmann[1,2] & Andrés Chavarría-Krauser[1,2*]

[1] Center for Modelling and Simulation in the Biosciences, Universität Heidelberg, Im Neuenheimer Feld 267, 69120 Heidelberg, Germany

[2] Interdisciplinary Center for Scientific Computing, Universität Heidelberg, Im Neuenheimer Feld 368, 69120 Heidelberg, Germany

Running title: **Modelling zinc uptake and radial transport in roots**

* For correspondence. E-mail andres.chavarria@bioquant.uni-heidelberg.de

# ABSTRACT

- *Background and Aims* Zinc uptake in roots is believed to be mediated by ZIP (ZRT-, IRT-like Proteins) transporters. Once inside the symplast, zinc is transported to the pericycle, where it exits by means of HMA (Heavy Metal ATPase) transporters. The combination of symplastic transport and spatial separation of influx and efflux produces a pattern in which zinc accumulates in the pericycle. Here, mathematical modelling was employed to study the importance of ZIP regulation, HMA abundance and symplastic transport in creation of the radial pattern of zinc in primary roots of *Arabidopsis thaliana*.

- *Methods* A comprehensive one-dimensional dynamic model of radial zinc transport in roots was developed and used to conduct simulations. The model accounts for the structure of the root consisting of symplast and apoplast and includes effects of water flow, diffusion, and cross-membrane transport via transporters. It also incorporates the radial geometry and varying porosity of root tissues, as well as regulation of ZIP transporters.

- *Key Results* Steady state patterns were calculated for various zinc concentrations in the medium, water influx and HMA abundance. The experimentally observed zinc gradient was reproduced very well. Increase of HMA or decrease in water influx led to loss of the gradient. The dynamic behaviour for a change in medium concentration and water influx was also simulated showing short adaptation times in the range of seconds to minutes. Slowing down regulation led to oscillations in expression levels, suggesting the need for rapid regulation and existence of buffering agents.

- *Conclusions* The model captures the experimental findings very well and confirms the hypothesis that low abundance of HMA4 produces a radial gradient in zinc concentration. Surprisingly, transpiration was found to be also a key parameter. The model suggests that ZIP regulation takes place on a comparable time scale as symplastic transport.

# INTRODUCTION

Zinc is a heavy metal and essential micronutrient for the growth of higher plants (Sommer and Lipman, 1926). As part of the functional subunits or cofactor of more than 300 proteins, among them the zinc-finger-proteins and RNA-polymerases, it plays an important role in many cellular processes. There are reports of zinc ions protecting plant cells from oxidative stress mediated by reactive oxygen species (ROS) (Cakmak, 2000) and they may act as an intracellular second messenger (Yamasaki et al., 2007).Hence, zinc deficiency is a wide-spread problem. Elevated zinc concentrations occurring especially on contaminated soils, e.g. in mining or industrial areas, lead to toxicity symptoms such as reduced growth and leaf chlorosis in most plants (Broadley at al., 2007; White, 2012). Some specialized zinc-hyperaccumulating species, however, are able to tolerate high levels of zinc without any detrimental effects (Zhao et al., 2000). These species possess mechanisms for both the increased uptake of zinc from the soil and its sequestration and detoxification (Macnair et al., 1999). These mechanisms implicate interesting applications in phytoremediation or nutritional enhancement (Chaney et al., 1997; Clemens et al., 2002) and have therefore been in the focus of scientific research. Through the soil zinc is supplied to the plant roots in the form of a divalent cation in aqueous solution. Mobilization by secretion of chelators and acidification of the rhizosphere might be necessary to dissolve zinc from soil particles prior to uptake (Clemens et al., 2002). Unlike water, charged zinc ions are virtually unable to cross cell membranes freely (Alberts et al., 2007) and therefore need to be transported into the root cells by specialized transporter proteins (Guerinot, 2000; Clemens et al., 2002). These transporter proteins are tightly regulated in order to provide a sufficient intracellular zinc concentration without reaching toxicity.

*Radial transport in the root*

Water and micronutrients like zinc are taken up from the soil by root cells and transported radially towards the xylem, from where they are distributed to stem and leaves (Clemens et al., 2002). These substances need to pass through several tissues: the epidermis, the cortex, the endodermis, and the pericycle (Hanikenne et al., 2008; Fig. 1). In *Arabidopsis thaliana*, each of these tissues comprises only one layer of cells (Dolan et al., 1993). The cytoplasm of adjacent cells is connected by cytoplasmic bridges in the cell wall, the plasmodesmata, which may be simple channels or have complex geometries (Roberts and Oparka, 2003). Thus, the cytoplasms of neighbouring cells form a symplastic continuum without membrane barriers.

In addition to the symplast, water and ions can also move in the cell wall, the apoplast, which has been found to contribute significantly to root transport processes (Steudle, 1994). The apoplastic flow, however, is interrupted in the endodermis by suberin deposited in the cell wall (Casparian strip). This strip is mostly impermeable to water and ions, although some findings suggest there may also be flow across this barrier (White et al., 2002; Ranathunge et al., 2005). Nevertheless, most water and ions need to pass the cell membrane before the Casparian strip to be transported further in the symplastic pathway (Yang and Jie, 2005). Because membrane transport is much more selective than the apoplastic flow, this barrier is believed to function as a mechanism to control the selectivity and magnitude of nutrient delivery to the xylem. While considerable amounts of water can cross the membrane freely, ions are almost completely blocked and their transport across cell membranes relies on specialized transporter proteins (Clemens et al., 2002).

*Zinc transporters*

There are different transporter proteins involved in the transport of zinc in plants. The most well-known families are: the ZIP (ZRT-, IRT-like Proteins) family, the HMA (Heavy Metal ATPases) family, and the MTP (Metal Tolerance Protein) or CDF (Cation Diffusion Facilitator) family. It appears that members of the ZIP family are responsible for zinc influx into the cytosol, HMAs for the efflux of zinc into the apoplast (needed for xylem loading), and MTPs for the sequestration of zinc to intracellular compartments such as the vacuole (Palmer and Guerinot, 2009). Other results suggest the involvement of YSL (yellow-stripe-like) transporters and OPT (oligopeptide transporters) in zinc homeostasis and transport of chelated metal ions (Schaaf et al., 2005; DiDonato et al., 2004).

ZIP family transporters accomplish the influx of zinc into root cells. The main transporters situated in the cell membrane and involved in zinc transport appear to be ZIP1 to ZIP6, ZIP9, ZIP10, ZIP12 and IRT3 (Grotz et al., 1998; Guerinot, 2000). Some of these transporters are expressed at very high levels under conditions of zinc deficiency, whereas their expression strongly decreases within less than two hours when zinc is added to the surrounding medium (Talke et al., 2006; Mortel et al., 2006). In addition, iron-regulated transporters IRT1 and IRT2 may be involved in zinc transport (Shanmugam et al., 2011).

The exact mechanism of this regulation is mostly unknown, but recent results have shown that the *ZIP4* gene in *A. thaliana* is regulated by transcription factors of the basic-region leucine zipper (bZIP) family, namely bZIP19 and bZIP23 (Assunção et al., 2010a). These factors bind to a ZDRE (Zinc Deficiency Response Element), which can also be found in the upstream regions of the genes of *ZIP1, ZIP3,* and *IRT3*.

How the bZIP19 and bZIP23 transcription factors sense internal zinc concentrations is unclear, since they do not appear to have a zinc binding site. It has been proposed that there are further players able to bind zinc and act as inhibitors of bZIP19 and bZIP23 (Assunção et al., 2010b). It has been found in other regulatory networks with bZIP transcription factors that they can be regulated post-transcriptionally in various ways, e.g. by protein binding or phosphorylation (Schütze et al., 2008). Very often, these factors act as dimers (Jakoby et al., 2002). bZIP19 and bZIP23 seem to be partially redundant (Assunção et al., 2010a) and it is supposed that they preferentially form homodimers, but may also interact to constitute heterodimers (Deppmann et al., 2006).

The efflux of zinc from the root to the shoot mainly depends on HMA2 and HMA4 transporters, which are predominantly expressed in the pericycle cells adjacent to the xylem (Sinclair et al., 2007; Hanikenne et al., 2008). Zinc hyperaccumulator species like *Arabidopsis halleri* appear to have the same ZIP transporters as non-hyperaccumulators, but different *HMA4* genes. Moreover, studies in different plant species have shown that hyperaccumulators possess multiple copies of *HMA4* in their genome. This results in higher expression levels and more efficient root-to-shoot transport of zinc (Hanikenne et al., 2008; Ó Lochlainn et al., 2011).

*Models of water and solute transport*

There are several approaches to model water and solute transport through root tissues. One approach is to model transport in analogy to electric resistor networks according to Ohm's and Kirchhoff's laws (Steudle and Frensch, 1996; Steudle and Peterson, 1998). Katou and Taura (1989) and Taura et al. (1988) use advection-diffusion equations to describe water and solute movement in the apoplast. Many modelling approaches concern the interface between soil and root surface (Ptashnyk et al., 2011; Leitner et al., 2009; Zygalakis et al., 2011). However, to the knowledge of the authors, there has so far been no attempt to couple a structured transport model in the root tissue to a regulatory model for transporters.

The regulatory mechanism of ZIP transporters has been modeled in a recent paper (Claus and Chavarría-Krauser, 2012). For the sake of simplicity, the root was modelled as one compartment in that publication, disregarding the fact that uptake, symplastic transport and xylem loading involve several different cell types (Clemens et al., 2002; Hanikenne et al., 2008). This paper extends the original model to consider the internal structure of root tissues in more detail. Symplastic and apoplastic transport in a radial geometry is coupled to the regulatory mechanism to understand the accumulation pattern of symplastic zinc and to find the prerequisites of moving zinc ions from the root surface to the xylem.

# METHODS

A modelling approach was used to conduct computer simulations of radial zinc transport in *A. thaliana* roots. The model consists of a coupled system of ordinary differential equations describing the regulation of ZIP transporters for each cell and 1-D partial differential equations describing the spatio-temporal evolution of concentration in the symplast and apoplast. Only a short description of the model is given below. The interested reader is referred to the supplemental file for a detailed derivation.

*Assumptions*

The root geometry was simplified as a single radially symmetric cylinder and transport in the root was assumed to take place in the radial direction only. This allowed reducing the three-dimensional problem into coupled one-dimensional problems in the later treatment. The structure of the root along the radius is shown schematically in Fig. 1. The root was assumed to be composed of the following cell types (from outside to inside): epidermis (*ep*), cortex (*co*), endodermis (*en*) and pericycle (*pc*). The cell layers extend from radius $r_x$ to $r_e$. A perfectly impermeable Casparian strip (*cs*) was assumed at position $r_c$ in the endodermis.

Roots express several different influx and efflux proteins. For the sake of simplicity, we assumed the existence of only two types of transporters: influx (ZIP) and efflux (HMA4). Epidermal, cortal and endodermal cells were allowed to have ZIP transporters, while pericycle cells produced only HMA4 transporters. Following the results of Talke et al. (2006), the expression of *HMA4* was assumed to be independent of the zinc concentration and was included into the model as a given amount of transporters. Transport across the membranes via ZIP and HMA4 was modelled as an enzymatic reaction with Michaelis-Menten kinetics. The model uses no other type of signal besides the internal zinc concentration. Hence, coordination is achieved merely by zinc fluxes.

Cells have a complex internal structure with organelles, such as vacuoles, nucleus, etc. They are also interconnected by plasmodesmata, which reduce the flow cross section substantially. To avoid the treatment of these internal structures, we regarded the cell content to be a porous medium with a given volume fraction. Vacuoles were considered only by a reduction of flow cross section, i.e. they were not treated as separate compartments and their role in sequestration was neglected. Cell walls were also assumed to be a porous medium of constant structure and porosity. We introduced a volume fraction for the symplast, which depended only on the radial position. This assumption is valid in view of the periodic structure of the root and the orientation of cell layers (Fig. 1). The volume fraction of the apoplast was assumed to be constant, and based on the results of Kramer et al. (2007) it was set to have a value of 1/15. Fig. 2 presents at the bottom graph the volume fraction of the symplast used in the simulations. The volume fraction in plasmodesmata is of the order of 0.15 (Rutschow et al., 2011), while the vacuole was assumed to make up 0.8 of the cell volume giving a volume fraction of 0.2. These parameters are coarse estimates, but the exact values are not of crucial importance to the model and the simulation results.

*Zinc transport*

Transport of zinc was modelled by diffusion in the cell wall and diffusion-advection in the symplast. Note that we assumed no water fluxes in the apoplast, i.e. advection in the apoplast was neglected. Based on mass conservation of zinc, three equations for the apoplastic and symplastic concentrations in three dimensions of space can be derived:

$\partial_t (\Psi Z) - \text{div}(\Psi D \text{ grad } Z) = 0$    in apoplast,    (1a)

$\partial_t (\Phi Z) + \text{div}(\Phi v Z - \Phi D \text{ grad } Z) = 0$    in symplast,    (1b)

where $\Psi$ is the volume fraction of the apoplast, $\Phi$ is the volume fraction of the symplast, $Z$ is the zinc concentration, v the water flow velocity and $D$ the diffusion coefficient.

Solving these equations would deliver the time evolution of three dimensional distributions of zinc in the root tissue. For this purpose, a precise 3-D representation of the tissue and computationally expensive numerical methods would be needed. To avoid this but still capture the essential features on the tissue structure shown in Fig. 1, we focused on the radial distribution by reducing Eqs. (1a,b) into a system of 1-D equations:

$$\partial_t (\Psi\, r\, Z) + \partial_r ((D/r)\, \Psi\, r\, Z - \Psi\, D\, \partial_r (r\, Z)) = r\, Q, \qquad \text{in apoplast,} \qquad (2a)$$

$$\partial_t (\Phi\, r\, Z) + \partial_r ((v + D/r)\, \Phi\, r\, Z - \Phi\, D\, \partial_r (r\, Z)) = r\, Q, \qquad \text{in symplast,} \qquad (2b)$$

Here, $r$ denotes the radial coordinate and $Q$ the membrane fluxes into and out of the respective compartments. These equations describe the time evolution of the radial distribution of zinc in the apoplast and in the symplast, and were used to conduct the simulations.

In addition to diffusion and advection, zinc fluxes through the membrane have to be considered (ZIP and HMA4 transporters). These fluxes are modelled as chemical reactions taking place on the membrane

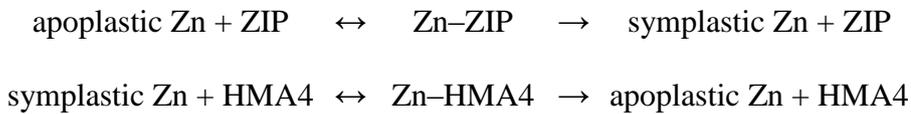

$$\text{apoplastic Zn} + \text{ZIP} \;\leftrightarrow\; \text{Zn–ZIP} \;\rightarrow\; \text{symplastic Zn} + \text{ZIP}$$
$$\text{symplastic Zn} + \text{HMA4} \;\leftrightarrow\; \text{Zn–HMA4} \;\rightarrow\; \text{apoplastic Zn} + \text{HMA4}$$

The first reaction describes uptake by ZIP, while the second reaction describes efflux by HMA4. Because the amount of free transporters is finite, the mechanism must saturate. As proposed by Claus and Chavarría-Krauser (2012), saturation was modelled by a Michaelis-Menten function instead of considering the full reactions given above. Influx was assumed to take place on the surfaces of epidermal, cortal and endodermal cells, while efflux was assumed to take place only at pericycle cells. This mechanism was coupled to the zinc conservation equations Eq. (2a,b) via a transporter dependent zinc flux density [compare supplemental Eqs. (S12) and (S18)]. More details on the transport model and its derivation are found in the supplemental file.

*Water flow*

As described above, zinc is carried along the water flow path with the velocity of the water (advection). This process influences the distribution of zinc and determines how fast variations in external zinc concentration spread in the system. To avoid a complete treatment of water fluxes in root tissues, we focused only on mass conservation delivering the radial flow speed by considering the effective flow cross section. Variation of cross section in the symplast was included by the volume fraction shown in the bottom graph of Fig. 2. Water fluxes in the apoplast were assumed to be small and were neglected, although the apoplast is believed to contribute to the total flux (Steudle, 2000). Hence, water exchange was assumed to occur only at the medium-epidermis (especially root hairs) and pericycle-xylem interfaces. Epidermal cells were assumed to take up water from the medium with a given constant flux velocity $q_0$. To simulate different transpiration rates the value of $q_0$ was varied. This approach is very simplistic. More sophisticated water flux models have been proposed by other authors, for example Katou and Furumoto (1986), Katou et al. (1987), Taura et al. (1988), Katou and Taura (1989), and Murphy (2000). These models, however, do not couple water flow to regulation of membrane transporters.

A consequence of mass conservation, incompressibility of water, and the radially oriented flow is that the mass flux through any two concentric surfaces must be equal. As the cross section decreases towards the stele, the flow speed has to increase proportionally to compensate for the smaller cross section (see Fig. S.3 in supplemental file). The top graph in Fig. 2 presents the flow velocity used in the sequel. It shows clearly how the cylindrical geometry of the root results in a general increase of velocity towards the stele. Note that the flow profile can be computed independently from the zinc concentration.

*Regulation*

Regulation of ZIP transporters has been discussed in detail in a previous paper (Claus and Chavarría-Krauser, 2012), where we found an activator-inhibitor model with dimerisation to present a likely mechanism. In this paper, steady state parameters were fitted with an optimisation method. This model along with the fitted parameters was used here to describe the amount of ZIP transporters for each cell type shown in Fig. 1 except for the pericycle. A schematic diagram of the mechanism is presented in Fig. 3. An inhibitor *I* senses the internal zinc concentration and inhibits a dimerising activator *A*. Gene expression for the production of transporters depends on activation by *A*. An increase in internal zinc results in higher levels of inhibitor, lower levels of activator, lower production of transporter and, thus, less zinc influx – and vice versa for a decrease in concentration. More details on the regulation model are given in the supplemental file.

Although this regulatory mechanism was developed as an average description over the whole root, it is more realistic to assume individual gene expression and transporter production in each single cell. These regulatory systems had to be coupled to transport in the apoplast and symplast [Eq. (2a,b)], which deserves a few considerations. First, zinc is not homogeneously distributed within each cell and it is unclear where and how the cell actually senses the concentration. Here, we assumed that a cell senses its average cytoplasmic concentration. Depending on this sensed concentration, each cell adjusts its own expression level and the resulting amount of transporters independently of the other cells. The transporter proteins are assumed to be confined to the cell they are produced in and evenly distributed on the plasma-membrane within each single cell. This assumption is supported by the HMA2 expression pattern found in *A. thaliana* (Sinclair et al., 2007: Fig. 1(c) of that publication). The reader is referred to section S.2 of the supplemental file for a precise mathematical formulation of the coupling between transporter regulation and zinc transport.

*Parameters*

The diffusivity of zinc ions in water has been measured by Harned and Hudson (1952) to be 700 µm$^2$ s$^{-1}$. Values of 530 µm$^2$ s$^{-1}$ have been found for calcium in the axoplasm of Myxicola (Donahue and Abercombie, 1987). Based on these values we assumed an approximate diffusivity of $D = 500$ µm$^2$ s$^{-1}$ for zinc in the symplasm. This value does not consider the diffusion of chelated zinc, which could be several times slower. Diffusion in the cell wall has been measured to be fifteen times slower than in the cytoplasm (Kramer et al., 2007). The reduction in diffusivity was accounted for in the model by inclusion of a volume fraction of 1/15 [compare Eq. (2a,b)]. The order of magnitude of the surface water flux velocity $q_0$ was estimated based on data from Rosene (1943) for water fluxes through root hair cells and data from Zarebanadkouki et al. (2012) for *Lupinus albus* roots. To obtain similar steady state patterns as observed in fluorescence images of zinc distributions in roots (Sinclair et al., 2007; Hanikenne et al., 2008) $q_0$ was then manually adjusted within this range to values between 1 and 4 µm s$^{-1}$. The other unknown parameters $H_0$ and $T_0$ were also manually adjusted to qualitatively reproduce those data. The steady state parameters of the regulation model were adopted from Claus and Chavarría-Krauser (2012), where an automatic fitting procedure was used to obtain them from measured expression levels. Sizes of cells and tissue layers were estimated from Hanikenne et al. (2008) with $r_e = 40$ µm, $r_c = 12.5$ µm, and $r_x = 6$ µm. The most important model parameters are listed in Table 1 and the cell sizes used are listed in Table 2. Other parameters are found in Tables S.1 to S.3 in the supplemental file.

*Numerical methods*

A conservative (i.e. mass-preserving) finite difference method on an equidistant grid was applied to solve the system of partial differential equations in one space dimension [Eq. (2a,b)]. Operator splitting was used to employ different stable explicit finite difference schemes to the advective and diffusive contributions. For the advection term, a second-order McCormack method was applied. A first-order FTCS (Forward Time Centered Space) scheme was used for the diffusion and reaction terms. The boundary conditions were implemented with upwinding. To guarantee mass conservation, the boundary flux was corrected for the numerical diffusion of the scheme. The regulation model Eq. (S1) was solved by an explicit Euler method. Using explicit schemes for both systems allowed to couple the solvers without much effort. Fig. S.2 in the supplemental file illustrates the algorithm used to couple the different numerical schemes. Steady states were calculated with Newton's method using numerical derivatives. Average gradients from spatial data were obtained by linear regression.

# RESULTS

*Steady state*

Figure 4 shows the simulated steady state patterns of ZIP levels (boxes), and the zinc concentrations in the symplast (solid line) and apoplast (dashed line) for roots grown in media with high ($Z^e = 25$ µM) and low zinc ($Z^e = 1$ µM). In both cases, the apoplastic zinc concentration decreased from the epidermis towards the Casparian strip, as a consequence of uptake by ZIP, although this effect was clearer for high zinc. Behind the Casparian strip, the apoplastic concentration rose again due to HMA4 mediated efflux from the symplasm. The apoplastic concentration inside the stele was smaller for low external zinc. However, it was still about 60% of the concentration at high zinc, although the external concentration was 25 times smaller. The concentration in the xylem was very low in both cases, because inflowing water diluted the solution strongly. Regarding the symplastic concentration, a continuous gradient towards the stele was found. This gradient was substantially less obvious at low zinc, while strongly pronounced at high zinc. ZIP transporters were almost completely down-regulated in all cells for high zinc, while their expression was much higher for low zinc. The model predicts a gradient in ZIP activity with high activity in the epidermis and low activity in the endodermis. Although this gradient was still present for high zinc, the difference in expression level was very small and is probably hard to detect in a measurement.

*Variation of HMA4 and water influx velocity*

To model *A. thaliana* HMA4 mutants, simulations were performed with higher levels of HMA4 transporters (via parameter $H_0$; Table 1). As a reference value for the wild type we used $H_0=5$ µM µm$^{-2}$ s$^{-1}$. Figure 5 presents in the upper panel the symplastic concentration distribution for different HMA4 levels. Increasing HMA4 to two (dashed line) or three times (dotted line) the original level (solid line) led to a decrease of the overall zinc concentration and loss of the gradient. Surprisingly, a very similar behaviour was seen when varying the influx velocity of water (parameter $q_0$; Table 1).

Simulations with half (dashed line) or one fourth (dotted line) of the original velocity (see Figure 5, lower panel) showed a loss of symplastic zinc gradient similar to increasing HMA4. The overall zinc concentration, however, remained higher than in the variation of HMA4, with even slightly higher values in the epidermis than for the original velocity ($q_0=4$ µm s$^{-1}$). Very low water fluxes and very high HMA4 levels produced even a retrograde gradient with higher zinc concentrations in the epidermis than in the pericycle (data not shown).

To further investigate the relation between HMA4 level and symplastic gradient, the correlation between steady state average gradient and HMA4 activity was analysed. The average gradient was calculated by linear regression as the slope of a fitted linear function. In the case of an accumulation in the pericycle the result is a negative number, reflecting the fact that the gradient points inwards, i.e. concentrations increase towards the xylem. The absolute value of this gradient gives a measure of the "steepness" of the concentration profile. Figure 6 shows the dependence of the average gradient on the normalized HMA4 activity (solid line) and on the external zinc concentration (dashed line). While the response to variation of external concentration was fairly linear with an increase in steepness following an increase in concentration, the response to HMA4 was strongly non-linear. Raising HMA4 level led to a strong decrease in the absolute value of the gradient. For HMA4 levels larger than three times the wild type level, the gradient became even retrograde (change in sign). At less than half the wild type value, a singularity/pole was found. The gradient became very steep when approaching the pole and no steady state can be sustained beyond. The reason is an imbalance of influx and efflux, leading to more zinc being pumped into the symplast than can flow out into the xylem. Small variation of HMA4 produced a strong reaction near the pole, i.e. zinc accumulation reacted very sensitive to HMA4 activity there.

*Dynamics*

The temporal evolution of the response of the system to changes in environmental conditions was simulated to understand the transient behaviour of the interplay between regulation and transport.

Starting from the steady state at low zinc ($Z^e$ = 1 µM), the external zinc level was raised to a high zinc condition ($Z^e$ = 10 µM). This corresponds to a resupply experiment. Figure 7 presents the evolution of the concentration profiles in time during this adaptation. After increasing the external concentration zinc diffused quickly into the apoplast outside the Casparian strip (in less than 1 s). During this period, regulation kept the high expression level of ZIP, resulting in an overall increase in symplastic concentration and a more pronounced gradient. This led to an increase in apoplastic concentration inside the stele and later to a sudden down-regulation of ZIP. Although the time scale of regulation was assumed to be equal to that of transport (parameter $\tau$ = 1; Table 1), adaptation of ZIP activity lagged behind, leading to an "overshoot" at about 5 s, where symplastic zinc in the pericycle exceeded its final steady state value by a factor of two. From there on the system stabilized to finally approach the new steady state value after about 20 s.

Transpiration, which defines the velocity of water flow, can vary substantially in the course of the day and is minimal during the night. To understand how adaptation to a change in transpiration rate takes place, we simulated the temporal changes upon a sudden change in water influx velocity $q_0$ from 0.05 to 4 µm s$^{-1}$ (see Figure 8). Due to the low water flux, a slightly retrograde concentration gradient was found at 0 s. ZIP expression was also almost constant in space, with a slightly higher expression in the endodermis. After increasing $q_0$, the concentration gradient built up quickly. An accumulation in the pericylce was clearly visible after 0.2 s and stabilized within 1 s. This led to a reversal of the distribution of gene expression, with a clear gradient towards the epidermis, where expression stabilized at twice the level of endodermis cells. The apoplastic concentration outside the Casparian strip did not change much during equilibration. In contrast, the apoplastic concentration in the stele decreased to half the initial value as a consequence of the higher water flow rate. For the same reason, the concentration in the xylem fell from about 10 µM for low transpiration to almost 0 µM for a normal transpiration rate.

*Time scale of regulation*

In the numerical experiments described above, the time scale of regulation was set to be comparable to the time scale of transport and diffusion (parameter $\tau = 1$; Table 1). The real time scale of regulation is unknown and may indeed be much slower. To understand how the specific choice of $\tau$ influences the entire process, simulations with different time scales were performed. Figure 9 shows the effect of resupply from low zinc ($Z^e = 1$ µM; upper black square) to high zinc ($Z^e = 10$ µM; lower black square) by plotting: the evolution in time of average ZIP activity and average internal zinc concentration as a phase diagram (Fig. 9A), and the average internal concentration against time (Fig. 9B). The paths shown in Fig. 9A represent the state of the root in time and the arrows mark the direction in which the state moved. The same transition from low to high zinc was conducted for three different time scale factors $\tau$: 1 (red path), 0.1 (blue path) and 0.01 (green path). This means that the time scale of regulation was approximately 10 s for $\tau = 1$, 100 s for $\tau = 0.1$ or 1000 s for $\tau = 0.01$.

In general, the internal concentration rose strongly after resupply, exceeding the one of the new steady state at high zinc ("overshoot", Fig. 9A, B). The reason was the high ZIP expression level at the initial state (about 30% activity; Fig. 9A). Regulation then reacted by shutting down the expression of ZIP (vertical portion of the paths; Fig. 9A), overreacting even slightly. For $\tau = 1$ the system eventually approached the new steady state. For slow regulation ($\tau = 0.1$ and 0.01) stable oscillations around the steady state were observed (Fig. 9A, B). In those cases the system did not approach the steady state. The reason for the oscillation was the overreaction of regulation producing a strong sudden reduction in concentration (horizontal path; Fig. 9A), which led again to an overreaction of upregulation, and to a too large concentration. The amplitude of the oscillation correlated with the time scale of the regulation: slow regulation produced larger oscillations (Fig. 9A, B).

# DISCUSSION

*Steady state*

Fluorescence imaging data (Sinclair et al., 2007; Hanikenne et al., 2008) show higher zinc concentrations in the apoplast than in the symplast. Our simulations reproduce this behaviour, although the accumulation in the apoplast does not appear as prominent as in the fluorescence images. This may be explained by several factors that lead to an underestimation of zinc in the symplast and an overestimation in the apoplast in the fluorescent images. Firstly, zinc and other cations can bind to components of the cell wall and accumulate in the apoplast (Sattelmacher, 2001), which is not considered in our model. Secondly, the fluorophore Zinpyr-1 used by Sinclair et al. (2007) and Hanikenne et al. (2008) reflects only levels of non-chelated zinc (Sinclair et al., 2007). Considering, in addition, that the vacuole contributes up to 90% of the cell volume, the fluorescence images reflect rather the concentration in the vacuole.

Within the symplast, a radial concentration gradient with accumulation in the pericycle has been found in experiments (Sinclair et al., 2007; Hanikenne et al., 2008). This pattern was reproduced very well by the model under the following conditions. We found that for existence of the pattern the contributions of influx, efflux, diffusion and advection have to be properly balanced. Influx transporter activity via ZIP (parameter $T_0$; Table 1) needs to be about 100 times higher than efflux transporter activity via HMA4 (parameter $H_0$; Table 1) to sustain the pattern in equilibrium. The absolute values of $H_0$ and $T_0$ are not accessible to direct interpretation, but their ratio should give a good estimate for a real root. Much less HMA4 than ZIPs was needed to obtain the pattern, which could explain why roots express so many different ZIPs compared to HMAs. Higher influx or higher efflux produced retrograde gradients, when advection was not increased correspondingly. This result is surprising in the sense that the pattern can be expected to vary in the course of the day, since advection varies in roots as a consequence of changes in transpiration rates. Diffusion is by its nature an equilibrating process, which seeks to even out any concentration gradient. Hence, if advection is small compared to diffusion, accumulation in the pericycle cannot occur. Crucial points for the formation of the pattern are water flow and geometry. First, the radially oriented advection driven by water uptake is the most likely process that is physically able to create accumulation at the pericycle. It links the spatially separated influx and efflux cells. Although these are also linked by diffusion, accumulation cannot be explained by that process. The importance of the velocity of water influx will be discussed below in more detail. Second, the cylindrical geometry supports the formation of the pattern, because the volume contracts towards smaller radii. This accelerates water on its path to the xylem and concentration increases faster at small radii for the same flux (less solvent volume), both helping to create a larger accumulation in the pericycle. Without this geometrical effect, the pattern was far less pronounced (data not shown). To produce a sufficiently pronounced gradient the influx velocity of water, $q_0$, needed to be larger than 1 µm s$^{-1}$ (Fig. 5 bottom). Zarebanadkouki et al. (2012) measured velocities of approximately 0.2 µm s$^{-1}$ in *L. albus* during the day. Although the lateral roots of *L. albus* have a substantially larger diameter than the primary root of *A. thaliana*, the order of magnitude is comparable to our values. Measurement of the flux velocity in *A. thaliana* roots and an extension of the water flow model would be desirable to be able to draw more precise conclusions.

Regarding the patterns of ZIP expression, our model predicts that ZIP activity follows conversely the pattern of symplastic concentration. The spatial pattern in gene expression is particularly clear for low zinc ($Z^e = 1$ µM), where the predicted gene expression level in the epidermis is five times higher than in the endodermis. In Claus and Chavarría-Krauser (2012) we showed that the regulation mechanism proposed is particularly robust for certain internal concentrations (Fig. 6A of that publication). In this "robust" range, gene expression can vary substantially without much effect on the internal zinc concentration. Our model suggests that at an external zinc concentration around 1µM differences in expression level between epidermis and endodermis cells are most pronounced. This is the external concentration at which an experimental validation of the expression pattern should be conducted to obtain the clearest results. Measurements of expression and protein levels of the ZIP4 orthologue ZNT1 in *Noccaea caerulescens* (Milner et al. 2012) support our model in showing a gradient with higher expression at the epidermis and lower expression in cortex and endodermis. However, their data also show high expression of *ZNT1* in the stele, which is most likely due to a putative role of ZNT1 in long-distance transport of zinc that may be regulated by mechanisms not considered in our model. Measurements of Birnbaum et al. (2003) are not in concord with our results. Using growth medium with 30 µM external zinc, these authors measured the expression of various genes in different root cell types using a high-throughput method. They found the expression levels of ZIP2 and ZIP4 to be minimal in the cortex and similarly high in the epidermis and endodermis. The measurements of Birnbaum et al. (2003) are surprising in the sense that they are not in line with the well documented symplastic gradient. Their results suggest that concentration of zinc is maximal in the cortex cells, where expression was found to be lowest. Neither our model nor the fluorescence images of Sinclair et al. (2007) and Hanikenne et al. (2008) support this. As mentioned above, an experimental validation of the expression patterns should be conducted at an external concentration for which a high range in expression is expected. At 30 µM external zinc, at which Birnbaum et al. (2003) conducted their measurements, expression can be expected to be very low in general and the dynamic range per se small (see for example Fig. 4 top). Small gradients in expression level are probably insignificant compared to the uncertainty of the measurement at that high concentration. Future experiments are needed to verify the gradient in ZIP activity predicted by the model. In addition, other ZIP transporters should be measured, since ZIP2 and ZIP4, like ZNT1 in *Noccaea caerulescens* may play other roles in metal transport (Milner et al., 2012) and may therefore not be related to the zinc concentration pattern.

*Variation of HMA4 and water influx velocity*

The level of HMA4 has been increased in experiments by introducing a *HMA4* gene of the zinc hyperaccumulator *A. halleri* into roots of *A. thaliana*, which leads to higher expression and more efficient transport of zinc into the xylem (Hanikenne et al., 2008). As a result, fluorescence images showed a change in the distribution of zinc in the tissue such that accumulation in the pericycle and the radial gradient were lost. Indeed, this effect is captured by the model very well, as increasing HMA4 to two or three times the original level led to a decrease of the overall zinc concentration and loss of the gradient (Fig. 5 top and Fig. 6). The contrary effect has been observed in *hma2, hma4* double mutants of *A. thaliana* (Sinclair et al., 2007) and in *A. halleri* with reduced expression of *AhHMA4* (Hanikenne et al., 2008). Our model describes this situation also very well (Fig. 6) and predicts a high sensitivity to variations in HMA4 level in this regime. Talke et al. (2006) showed that *HMA4* expression in *A. thaliana* and *A. halleri* roots varies substantially less in resupply and oversupply experiments than the one of *ZIP* genes. From this observation the authors concluded that its expression does not depend much on the zinc status. In view of a regime of high sensitivity predicted by our model, HMA4 might actually also be subject to regulation and only small adaptation of expression might be enough to create sufficient effect. This may explain why *HMA4* expression varies less.

Surprisingly, decreasing the influx velocity had a very similar effect as increasing HMA4 (Fig. 5). Since transpiration rate is largely determined by external factors such as the time of day, light and humidity the flow velocity of water in roots is highly variable. Simulations with half or one fourth of the original velocity showed a loss of the radial zinc gradient. This effect is similar to an increase in HMA4, yet the overall zinc concentration remained slightly higher. While an increase in HMA4 caused enhanced efflux of zinc into the xylem and thereby an increase in the total efflux from the symplastic domain, a decrease in velocity only changed the distribution of zinc in the tissue. Instead of accumulating zinc in the pericycle, diffusion dominated and produced an almost homogeneous distribution. Adaptation to a new transpiration rate is predicted to take place within less than a minute (Fig. 8). No published experimental data examining the relation between transpiration rate and zinc localization in roots is known to the authors. One conclusion that can be drawn from our model for future experiments is that transpiration rate and water status of the plant have to be controlled precisely to avoid artefacts, as just moving the plant into dark to conduct the measurement might change the pattern.

The model also predicts that external zinc concentrations influence the strength of the gradient in an almost linear manner (Fig. 6), with higher external concentrations leading to higher accumulation of zinc in the pericycle. No experimental quantifications of the concentration gradient for varying concentrations are known to the authors. Sinclair et al. (2007) found an almost linear relation between Zinpyr-1 fluorescence and zinc concentration in the medium, suggesting that the average internal concentration depends linearly on the external concentration. Our model also predicts a roughly linear dependency with a slope of approximately 0.09 µM µM$^{-1}$. Talke et al. (2006) also measured the average zinc concentration in *A. thaliana* and *A. halleri* roots for different external concentrations and found a positive correlation. They plotted their data with a logarithmic scale, which makes the direct comparison more difficult and error prone. However, plotting our linear relationship against a logarithmic scale produces a graph similar to the one published by Talke et al. (2006).

*Dynamics*

Similar to the resupply experiments by Talke et al. (2006), we simulated the adaptation of the system to a new environmental condition with increased external zinc. In our simulations, the new steady state was reached within 20 seconds (Fig. 7). Unfortunately, Talke et al. (2006) measured the expression levels only every two hours, which is too coarse to resolve the dynamics of adaptation in the root, as suggested by our results. Therefore, there is also no experimental validation of the overshoot seen in our simulations. Since adaptation of the regulatory system is relatively slow compared to transport, this "overshoot" was produced, where zinc concentration temporarily exceeded the final value before reaching the new steady state.

While a sudden change in the external zinc concentration is rather unlikely in a natural environment, changes in the flow velocity happen regularly as a consequence of changes in leaf transpiration rates. Therefore, this potentially frequent change in environmental conditions was simulated in the model and the simulations show that here the new steady state is reached within 5 seconds and that there is no overshoot as described above.

Equilibration predicted by our model seems at a first glance to take place very fast (within less than a minute). Although some of the parameters in the model have been chosen arbitrarily to match the accumulation pattern qualitatively, equilibration within minutes is plausible for the following reasons. The equilibration time is in principle determined by four parameters beside the diameter of the root (Table 1): $q_0$, $D$, $H_0$ and $T_0$. Here, $D$ and $q_0$ were estimated based on experimental findings, while $H_0$ and $T_0$ were chosen to qualitatively match the fluorescence images of Sinclair et al. (2007) and Hanikenne et al. (2008). Due to the qualitative nature of these images and the lack of measured average radial profiles, no automatic fitting procedure to obtain $H_0$ and $T_0$ could be applied, leaving some uncertainty in our predicted time scale. Nevertheless, the time scale is mostly fixed by the diffusion coefficient $D$, which determines the apoplastic transport and poses the most important process acting against accumulation by balancing symplastic concentration differences. The estimation of $D$ relied on measurements of free calcium in the axoplasm of Myxicola (Donahue and Abercombie, 1987). Since zinc is known to be chelated in the cytoplasm (Clemens et al., 2002), symplastic diffusion coefficients may be an order of magnitude lower, depending on the size of the chelator. The other parameters can be adapted to this situation by multiplying them with the same factor keeping the steady state accumulation pattern unchanged but rendering a slower equilibration. Assuming that the apparent diffusivity of chelated zinc is ten times lower than the value used here, the time predicted by our model would be roughly ten times higher and equilibration would take place within three to four minutes. Equilibration times of hours are not in concord with the time scale of diffusion in the symplast and apoplast. Therefore, our model indicates that experiments should be conceived to capture effects that live only a few minutes and not hours.

While the time scales of transport processes are substantiated to some extent by experimental findings, the time scale of ZIP regulation is largely unknown. Therefore, simulations with different time scales were performed to understand the origin of the overshoot found during equilibration. Variation of the relative time scale between transport and regulation showed that a slow regulation led to oscillations with high amplitudes (Fig. 9). Experimental validation of these oscillations may be difficult, because a real root lacks a high degree of spontaneous synchronisation and measurements are usually performed on tissue level rather than in single cells. The model assumes that the root is perfectly synchronised resulting in well-defined oscillations. In real roots, neighbouring cells may differ in phase and the oscillatory behaviour would rather come to light as a high variance in, for example, expression level. Since zinc shocks caused by oscillations can be toxic and even lethal to plant cells, regulation either needs to be fast – on the same time scale as transport – or the cells need to have fast and efficient buffering mechanisms. So far, our model neglects buffering and sequestration, although zinc is known to be sequestered into the vacuole (Clemens et al., 2002). To be efficient, these mechanisms would have to be fast and thus are unlikely to rely directly on gene regulation, which calls for a chelation mechanism. This point, however, needs to be treated in more detail in a future extension of the model.

*Conclusions*

Water uptake – and the associated advection of zinc towards the stele – is the main mechanism in formation of the radial zinc pattern in roots. The cylindrical geometry is a further factor supporting the effect of advection. The transpiration rate is expected to influence the pattern strongly and its reduction should produce similar effects to increasing the expression of *HMA4*. Zinc accumulation in the pericycle depends non-linearly on the expression of *HMA4*, where small variations of expression suffice to produce large effects. This might explain why HMA4 seems to be unchanged during resupply experiments. In general a much smaller activity of HMA4 than of ZIPs is needed to maintain sufficient zinc supply, giving a possible explanation why so many different zinc uptake transporters (ZIP family) are expressed in roots as compared to only few zinc release transporters (HMA). Resupply of zinc showed that regulation has to take effect within minutes to avoid strong peaks in symplastic concentration. Fast chelating agents seem to be necessary to dampen possible oscillatory behaviour and short-term oversupply. A slow sequestration possibly based on genetic regulation, such as one into the vacuole, is less suited to counteract these short-term effects and may be rather important in long term adaptation. We conclude that regulation has to be faster than assumed before.

## FUNDING

This work was funded by the German Research Foundation [grant number CH 958/1-1].

## ACKNOWLEDGEMENTS

The authors thank Ute Krämer, Ina Talke and Mohsen Zarebanadkouki for the fruitful discussions.

FIGURES

**Figure 1: Scheme of root tissues.** ZIP transporters are localized on the epidermis (ep), the cortex (co) and the endodermis (en). Endodermis cells are surrounded by the Casparian strip (cs). HMA4 transporters are localized in the pericycle (pc). The symbols $r_e$, $r_c$, and $r_x$ denote the radial coordinates at the outisde of the epidermis, the Casparian strip and the inside of the pericycle, respectively. Arrows illustrate the direction of transport.

**Figure 2: Flow velocity and volume fraction.** Relative flow velocity (top) and volume fraction (bottom) as functions of the radial coordinate. Tissues include the epidermis (ep), cortex (co), endodermis (en) and pericycle (pc).

**Figure 3: Scheme of ZIP regulation model.** The activator *A* dimerizes and induces gene activity, mRNA transcription and translation of the transporter protein. Zinc is transported across the cell membrane and causes the inhibitor *I* to inhibit *A*. Figure adapted from Claus and Chavarría-Krauser (2012).

**Figure 4: Spatial distribution of zinc in wild type.** Symplastic (solid line) and apoplastic (dotted line) zinc concentration for high (25 μM; top) and low (1 μM; bottom) medium concentration. In addition, grey boxes illustrate the ZIP activity in epidermis (ep), cortex (co) and endodermis (en).

**Figure 5: Variation of HMA4 and flow velocity.** Symplastic zinc concentrations are shown along the radial coordinate in epidermis (ep), cortex (co), endodermis (en) and pericycle (pc) cells. Top: HMA4 levels were increased to two (dashed line) or three times (dotted line) compared to the wild type (WT) level (solid line), while the flow velocity $q_0$ was kept constant at 4 μm s$^{-1}$. Bottom: Flow velocity $q_0$ was decreased from 4 μm s$^{-1}$ (solid line) to 2 μm s$^{-1}$ (dashed line) and 1 μm s$^{-1}$ (dotted line), while HMA4 was kept constant at the WT level.

**Figure 6: Symplastic zinc gradient.** Average symplastic zinc gradient shown as a function of HMA4 in multiples of the wild type levels (solid line) and of the external zinc concentration $Z^e$ (dashed line). Negative numbers reflect a gradient oriented towards the stele.

**Figure 7: Change in external zinc concentration.** Starting from the steady state distribution of zinc in the symplast (solid line) and apoplast (dashed line) at an external concentration of 1 μM, the medium concentration was changed to a high zinc condition (10 μM). Grey boxes show ZIP activity for epidermis (ep), cortex (co) and endodermis (en) cells. The new steady state is reached at $t = 20$ s.

**Figure 8: Change in flow velocity.** Starting from the steady state distribution of zinc in the symplast (solid line) and apoplast (dashed line) at $q_0 = 0.05$ μm s$^{-1}$, the flow velocity was increased to $q_0 = 4$ μm s$^{-1}$. Grey boxes show ZIP activity for epidermis (ep), cortex (co) and endodermis (en). A new steady state is reached after 5 s. External zinc was kept constant at $Z^e = 25$ μM.

**Figure 9: Variation of regulation time scale.** The time scaling factor $\tau$ of the ZIP regulation model was decreased from $\tau = 1$ (red line) to $\tau = 0.1$ (blue line) and $\tau = 0.01$ (green line) and time courses were simulated for a change of the external zinc concentration from 1 μM to 10 μM. A, the transition between the two steady states (black squares) is shown as a phase diagramm of ZIP activity again symplastic zinc. For $\tau = 1$, the system reaches the new steady state after a minor overshoot. For $\tau = 0.1$ and $\tau = 0.01$, the new steady is unstable and concentrations oscillate on a limit cycle. B, time course of the symplastic concentration showing clearly the oscillations.

TABLES

Table 1: List of parameters

| Parameter | Value | Description |
|---|---|---|
| $q_0$ | 0.1 – 4 µm s$^{-1}$ | Water flux velocity at root surface |
| $D$ | 500 µm$^2$ s$^{-1}$ | Diffusivity of zinc in cytoplasm |
| $r_x$ | 6 µm | Stele radius (without pericycle cells) |
| $r_c$ | 12.5 µm | Position of Casparian strip |
| $r_e$ | 40 µm | Root radius |
| $T_0$ | 500 µM µm$^{-2}$ s$^{-1}$ | Maximal amount of ZIP |
| $H_0$ | 5 µM µm$^{-2}$ s$^{-1}$ | Wild type amount of HMA4 |
| $\tau$ | 0.01 – 1 | Scaling factor for regulatory time scale |

Table 2: Cell sizes used in simulations. Values correspond to a typical young *Arabidopsis thaliana* root.

| Cell type | Left radius (µm) | Right radius (µm) |
|---|---|---|
| Pericycle | 6 | 9.75 |
| Endodermis | 10.25 | 14.75 |
| Cortex | 15.25 | 29.75 |

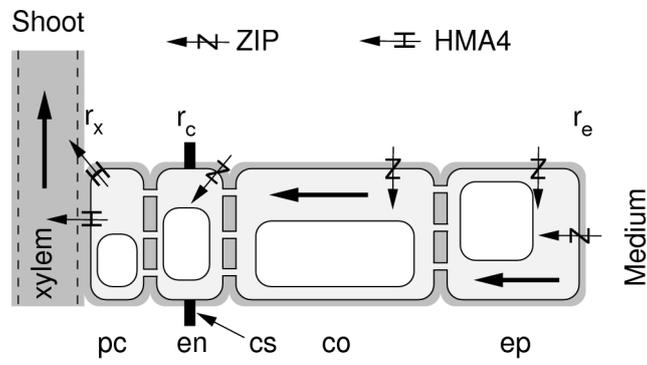

Figure 1:

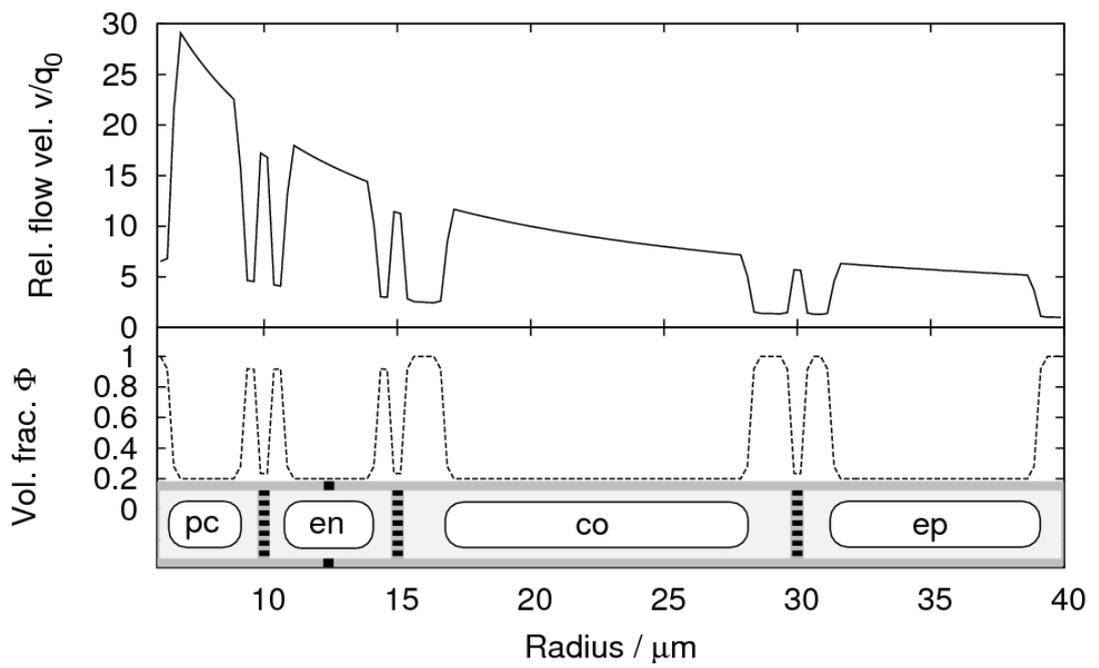

Figure 2:

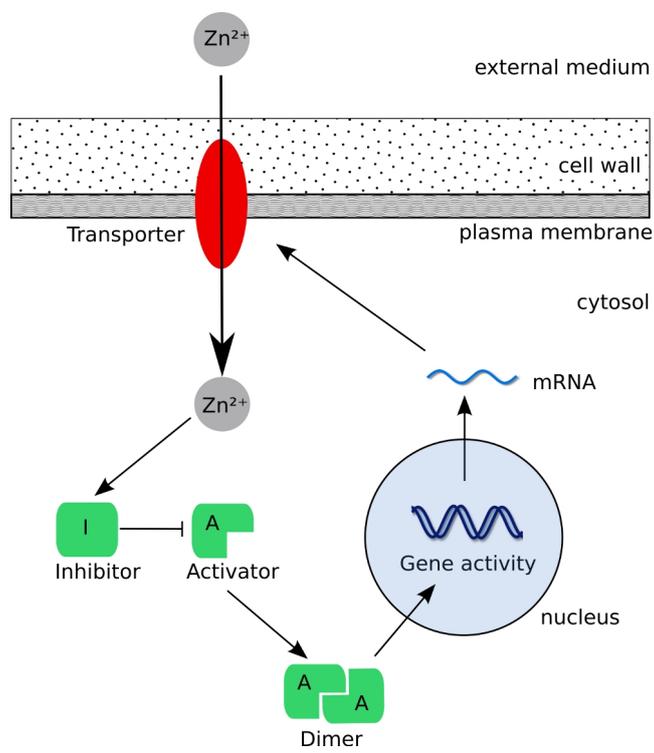

Figure 3:

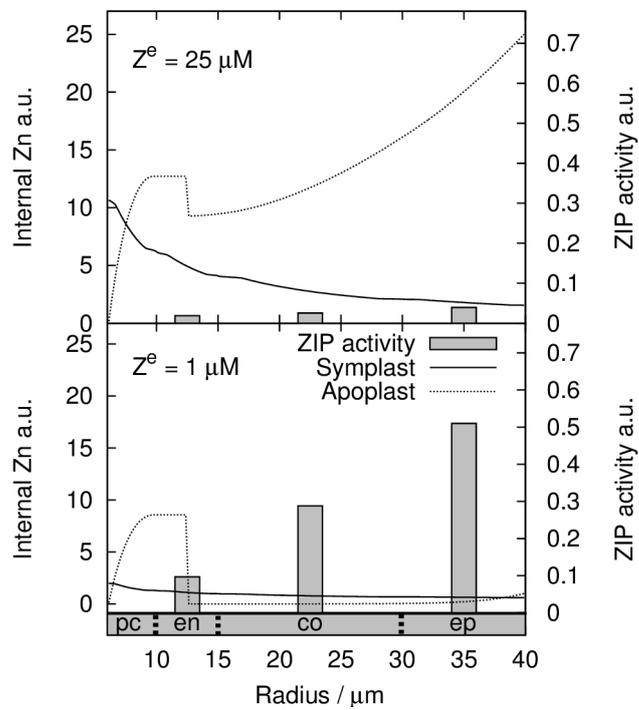

Figure 4:

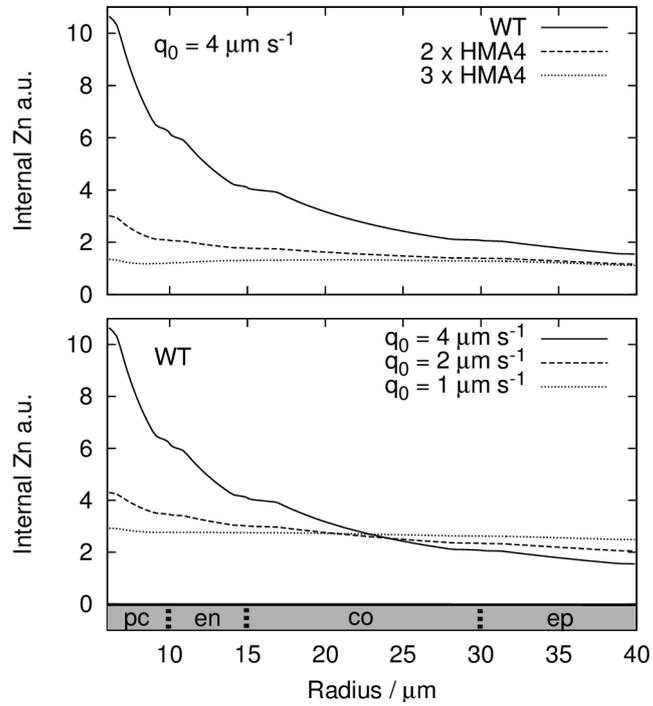

Figure 5:

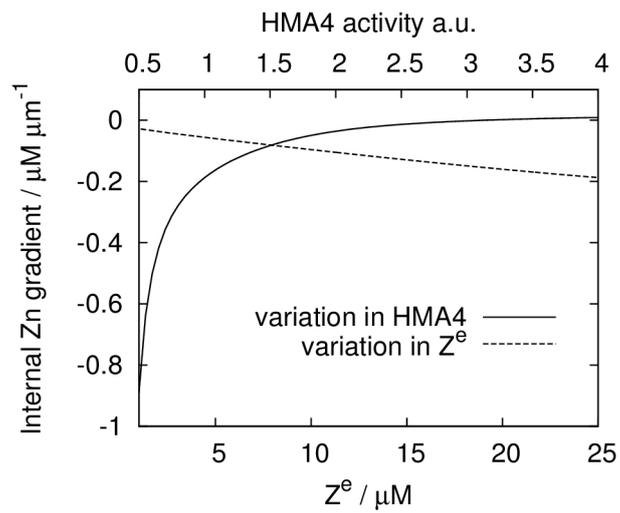

Figure 6:

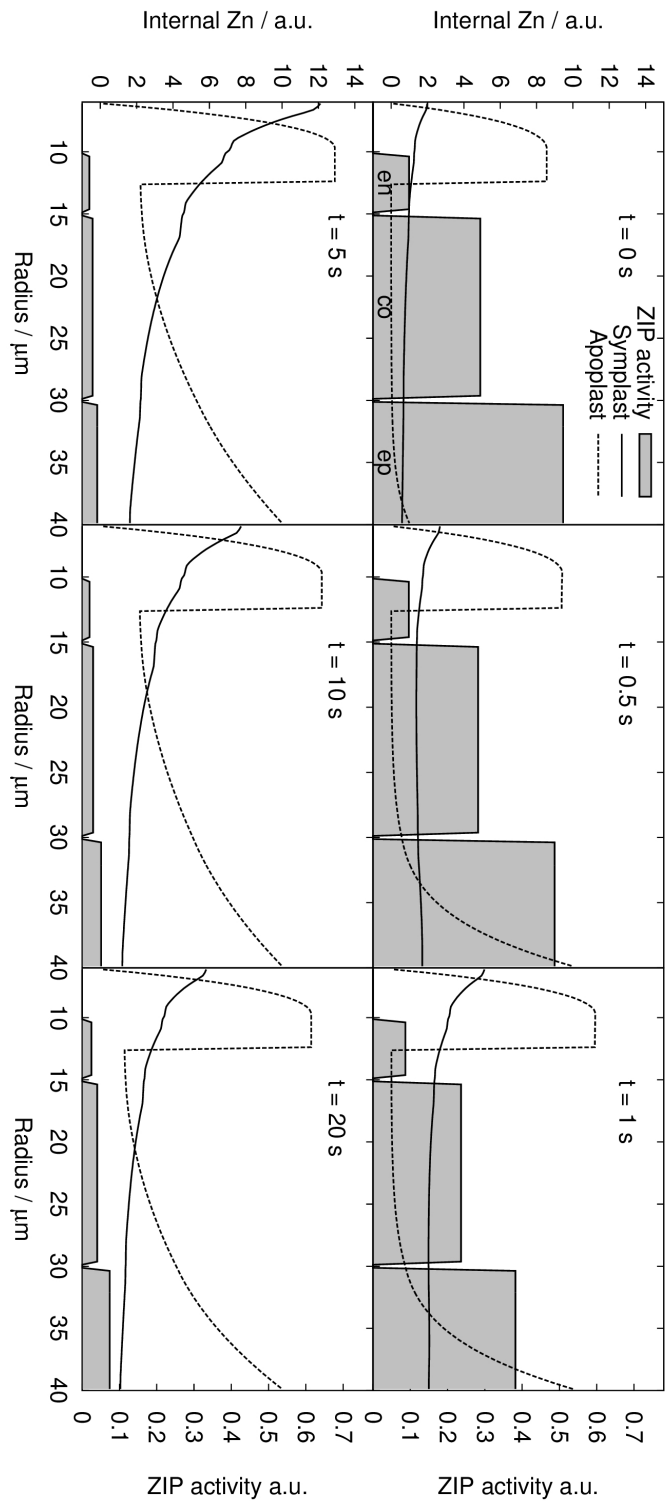

Figure 7:

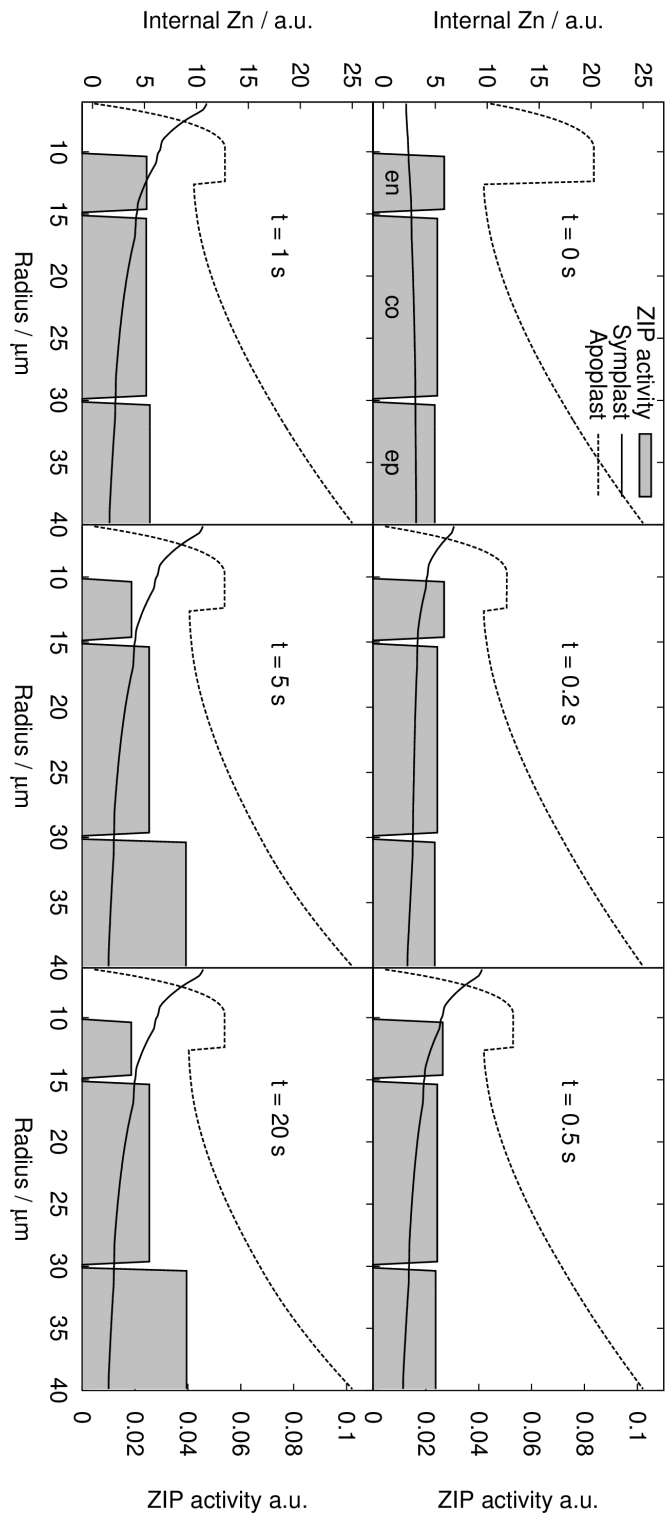

Figure 8:

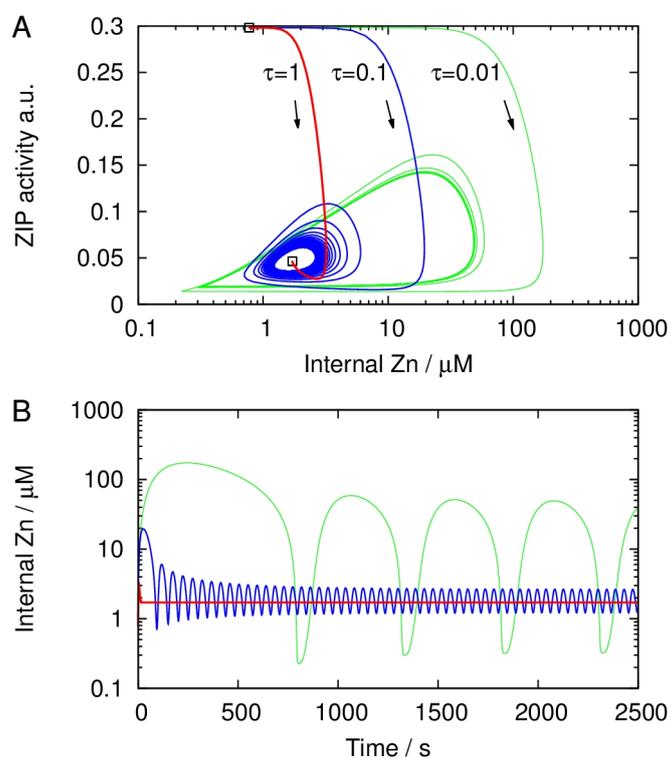

Figure 9:

# SUPPLEMENTARY DATA

## Zinc Uptake and Radial Transport in Roots of Arabidopsis thaliana: A Modelling Approach to Understand Accumulation

Juliane Claus[1,2], Ansgar Bohmann[1,2], and Andrés Chavarría-Krauser[*1,2]

[1]Center for Modelling and Simulation in the Biosciences, Universität Heidelberg, Heidelberg, Germany
[2]Interdisciplinary Center for Scientific Computing, Universität Heidelberg, Heidelberg, Germany

October 15, 2012

**Summary**

A cell-based model for the uptake and transport of zinc in roots of *Arabidopsis thaliana* and *Arabidopsis halleri* is derived. The model consists of a coupled system of ordinary differential equations describing the regulation of ZIP transporters and 1-D partial differential equations describing the transport in the symplast and apoplast. It considers, thus, the internal structure of the root tissue and couples transport phenomena with regulation networks. A system of ordinary differential equations for the xylem is also derived from a transport model and coupled via a boundary condition to the 1-D model of the apoplast.

## S.1 Overview

Transport in the root is assumed to be mostly in radial direction. This allows to reduce the three dimensional problem into coupled one dimensional radially oriented problems. The structure along the radius is shown schematically in Fig. 1 of the main manuscript. The root is assumed to be composed of following cell types (from outside to inside): epidermis (ep), cortex (co), endodermis (en) and pericycle (pc). The cell layers extend from radius $r_x$ to $r_e$. Surrounding the endodermis cells, a perfectly unpermeable Casparian strip (cs) at position $r_c$ is assumed. Epidermis, cortex and endodermis cells are assumed to possibly have ZIP transporters (influx), while pericycle cells have only HMA4 efflux transporters. Following the results of Talke et al. [2006], the expression of *HMA4* is assumed to be independent of the zinc concentration and will be included into the model as a given efflux. Both transporters ZIP and HMA4 are assumed to be saturable and to follow Michaelis-Menten kinetics. The expression of *ZIP* in the epidermis, cortex and endodermis is allowed to adapt to the current internal zinc status based on the dimerising activator-inhibitor model proposed by Claus and Chavarría-Krauser [2012]. Depending on the average internal zinc concentration, each cell adjusts independently of the others the expression level of *ZIP*. The resulting amount of transporters is assumed to be evenly distributed on the plasma-membrane and will vary in general from cell to cell (but not within). This assumption is supported by the HMA2 expression pattern found in *A. thaliana*, Sinclair et al. [2007] (Fig. 1(c) of that publication). Note that, the model needs no other type of signal beside the internal zinc concentration. Hence, coordination is achieved merely by zinc fluxes.

---

[*]andres.chavarria@bioquant.uni-heidelberg.de

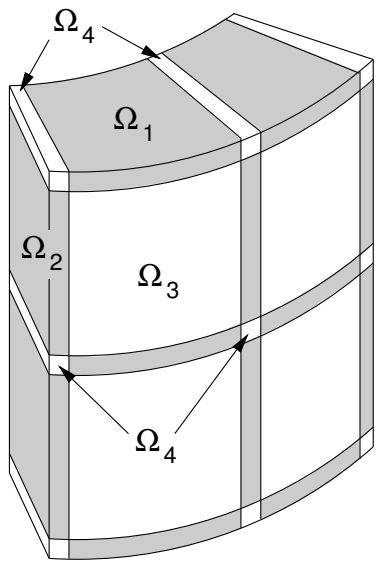

Figure S.1: Schematic drawing of the domains correspoding to apoplast $\Omega_1$, $\Omega_2$ and $\Omega_4$, and symplast $\Omega_3$. The root is assumed to be composed of a periodic assembly of such domains.

To be able to reduce the problem to a system of 1-D partial differential equations, some assumptions on the geometry are needed. First, the root is best described by cylindrical coordinates and reduction will be achieved by averaging over the axial coordinate $z$ and the azimuth $\varphi$. Second, assume that the domain of interest is made up by periodic repetition of $\bar{\Omega} := \bar{\Omega}_1 \cup \bar{\Omega}_2 \cup \bar{\Omega}_3 \cup \bar{\Omega}_4$ in azimuthal and axial direction. Here, $\Omega_3$ denotes the symplast and $\Omega_1$, $\Omega_2$, and $\Omega_4$ different parts of the apoplast respectively as indicated in Fig. S.1. And third, assume that the contribution of the apoplastic "edges" $\Omega_4$ to the overall zinc uptake is negligible. This assumption is justified, as $\Omega_4$ is thin compared to $\Omega_1$ to $\Omega_3$. Each layer of cells presented in Fig. 1 could be described by such a periodic assembly. We will assume that $\Omega_i$, for $i = 1, \ldots, 4$, span over all cell layers from $r_x$ to $r_e$.

Cells have a complex internal structure with organelles, such as vacoules, nucleus, etc. They are also interconnected by plasmodesmata – channels which traverse the cell wall. To avoid the treatment of these internal structures, we treat the cell content as a porous medium with a given volume fraction, i.e. the vacoules are not treated as separate structures. Therefore, we introduce a volume fraction $\Phi = \Phi(r)$ for the symplast, which depends only on the radial position. This assumption is valid in view of the periodic structure of the cell and orientation of cell layers (Fig. 1). Cell walls are also best described as a porous medium of constant structure and porosity with constant volume fraction $\Psi$.

## S.2 Regulation

Regulation is described by the dimerizing activator-inhibitor model proposed in Claus and Chavarría-Krauser [2012]. This leads to a system of five ordinary differential equations per cell type listed in Fig. 1

$$\left.\begin{aligned}
\frac{dG_\alpha}{dt} &= \gamma_G \left( K\, A_\alpha^2 \left(1 - G_\alpha\right) - G_\alpha \right), \\
\frac{dM_\alpha}{dt} &= \gamma_M \left( G_\alpha - M_\alpha \right), \\
\frac{dT_\alpha}{dt} &= \gamma_T \left( M_\alpha - T_\alpha \right), \\
\frac{dA_\alpha}{dt} &= \gamma_A \left( 1 - \Gamma\, A_\alpha\, I_\alpha - A_\alpha \right), \\
\frac{dI_\alpha}{dt} &= \gamma_I \left( \Gamma_I\, \zeta_\alpha - \Gamma'\, A_\alpha\, I_\alpha - (1 + \Gamma_I\, \zeta_\alpha)\, I_\alpha \right),
\end{aligned}\right\} \quad \begin{array}{l} \text{for } t \in (0, \infty), \\ \alpha = en, co, ep, \end{array} \quad \text{(S1)}$$

where $G_\alpha$ is the gene expression level, $M_\alpha$ the transcript level, $T_\alpha$ the transporter level, $A_\alpha$ an activator and $I_\alpha$ an inhibitor, and $\zeta_\alpha$ the internal zinc concentration. The factors $\gamma_i$, for $i = G, M, T, A, I$, are related to the time scales of each single reaction step. This system was non-dimensionalized in such a way that all variables take values between 0 and 1. Pericycle cells are not included, because these efflux cells can be assumed to not express an influx transporter. Fig. 3 in the main manuscript presents a diagram of the processes modelled by Eq. (S1).

The model expects one single value for the internal zinc concentration $\zeta_\alpha(t)$, while the transport model delivers a distribution $\mathcal{Z} = \mathcal{Z}(x, t)$ with $x \in \Omega_3$. Thus, the zinc concentration varies inside a single cell. To circumvent this issue, we assume that a cell senses the average zinc concentration inside it

$$\zeta_\alpha(t) = \zeta_0^{-1} \frac{1}{\mu(C_\alpha)} \int_{C_\alpha} \mathcal{Z}(x, t)\, dx \qquad \text{for } t \in [0, \infty),\ \alpha = en, co, ep, \quad \text{(S2)}$$

where $\zeta_0^{-1}$ is a scaling factor ($\zeta_\alpha$ is non-dimensionalized), integration is over the cell $C_\alpha \subset \Omega_3$ and $\mu(C_\alpha)$ is its volume. The transporters regulated by Eq. (S1) are assumed to be distributed evenly on the surface $\partial C_\alpha$ of the cell, and a distribution of transporters is constructed as follows

$$\mathcal{T}(x,t) = \sum_{\alpha \in \{en,co,ep\}} T_\alpha(t)\, \chi_{\partial C_\alpha}(x) \;, \qquad \text{for } (x,t) \in \overline{\Omega}_3 \times [0,\infty) \tag{S3}$$

with the characteristic function

$$\chi_{\partial C_\alpha}(x) = \begin{cases} 1 & \text{if } x \in \partial C_\alpha \;, \\ 0 & \text{if } x \notin \partial C_\alpha \;. \end{cases}$$

HMA4 efflux transporters at the pericycle are included in a similar manner

$$\mathcal{H}(x) = H_{pc}\, \chi_{\partial C_{pc}}(x) \;, \qquad x \in \overline{\Omega}_3 \;, \tag{S4}$$

where the assumption that the level $H_{pc}$ of HMA4 is constant was used, Talke et al. [2006].

## S.3 Water flow

Before treating our main topic of zinc transport in the root, we construct a simple model for water flow. Membranes restrict the movement of zinc, but water carries it along the flow path (advection). This process influences the distribution of zinc and determines how fast variations in external zinc concentration spread in the system. Ultimately, advection is essential to the regulation patterns.

To avoid a complete treatment of water fluxes in root tissues, we focus only on mass conservation delivering the flow speed by consideration of effective flow cross sections. Variation of cross section in the symplast is included through the volume fraction $\Phi(r)$. Water fluxes in the apoplast are assumed to be small compared to those in the symplast and will be neglected here. This includes the assumtpion that there is no exchange between symplast and apoplast, although the apoplast is believed to contribute to the total flux, Steudle [2000]. We assume that epidermal cells take up water from the medium with a constant flux density $q_0$. During its pathway to the xylem, water is assumed to be conserved and to have a flow speed that depends only on volume fraction and geometry (radial convergence). This approach is very simplistic and other more sophisticated models have been proposed, for example Katou and Furumoto [1986], Katou et al. [1987], Taura et al. [1988], Katou and Taura [1989], Murphy [2000]. Modeling water fluxes in plant tissues is a complex problem which deserves treatment of its own and is out of scope of this manuscript.

Mass conservation for an incompressible fluid reads in the symplast

$$\operatorname{div}(\Phi \mathbf{v}) = 0 \qquad \text{in } \Omega_3 \;, \tag{S5}$$

where $\Phi$ is the volume fraction and $\mathbf{v}$ is the flow velocity. Eq. (S5) is expressed in cylindrical coordinates $(r, \varphi, z)$ to reflect the geometry of roots

$$\frac{1}{r}\partial_r(r\,\Phi\,v_r) + \frac{1}{r}\partial_\varphi(\Phi\,v_\varphi) + \partial_z(\Phi\,v_z) = 0 \ . \tag{S6}$$

where $v_r$, $v_\varphi$, and $v_z$ denote the radial, azimuthal, and axial component of the velocity $\mathbf{v}$. To the end of reducing the model to 1-D we consider the surfaces

$$\Gamma_i(r) := \left\{\, \mathbf{x} \in \Omega_i \;\middle|\; x_1^2 + x_2^2 = r^2 \,\right\} \qquad \text{for } i = 1,\, 2,\, 3 \ , \tag{S7}$$

which can be described in cylindrical coordinates by

$$(r, \varphi, z) \in \{r\} \times (0, \varphi_{0,i}) \times (0, z_{0,i}) \qquad \text{for } i = 1,\, 2,\, 3 \ ,$$

with azimuth $\varphi_{0,i}$ and height $z_{0,i}$ of the considered domain. Note the use of different polar coordinate systems for $\Omega_1$, $\Omega_2$, and $\Omega_3$. Let $\mu_i(r) = r\,\varphi_{0,i}\,z_{0,i}$ denote the area of $\Gamma_i(r)$. Introduce the averaged radial velocity in the symplast $\Omega_3$

$$v(r) := \frac{1}{\mu_3(r)} \int_{\Gamma_3(r)} v_r(r, \varphi, z)\,\mathrm{d}\gamma \ , \qquad \text{for } r \in [r_x, r_e] \ ,$$

and consider the corresponding average of Eq. (S6) over $\Gamma_3(r)$:

$$\frac{1}{\mu_3(r)} \int_{\Gamma_3(r)} \frac{1}{r}\partial_r(r\,\Phi\,v_r)\,\mathrm{d}\gamma = -\frac{1}{\mu_3(r)} \int_{\Gamma_3(r)} \left(\frac{1}{r}\partial_\varphi(\Phi\,v_\varphi) + \partial_z(\Phi\,v_z)\right)\mathrm{d}\gamma \ ,$$

where the second and third terms in Eq. (S6) were moved to the right hand side. With the surface element $\mathrm{d}\gamma = r\,\mathrm{d}\varphi\,\mathrm{d}z$ the left hand side of the equation is

$$\frac{1}{\mu_3(r)} \int_{\Gamma_3(r)} \frac{1}{r}\partial_r(r\,\Phi\,v_r)\,\mathrm{d}\gamma = \frac{1}{r}\partial_r(r\,\Phi\,v) \ ,$$

while the terms on the right hand side correspond to a two dimensional divergence and can be transformed into an integral over the boundary $\partial\Gamma_i(r)$. This boundary integral is zero, based on the assumption that the apoplast and symplast do not exchange water (consequence of assuming no water fluxes in the apoplast). We obtain an equation for the average flow velocity in the symplast

$$\partial_r(r\,\Phi\,v) = 0 \qquad \text{for } r_x < r < r_e \ , \tag{S8}$$

$$\Phi v\big|_{r=r_e} = q_0 \ , \tag{S9}$$

where the water influx $q_0$ was prescribed on the root surface ($r = r_e$). This system can be solved by integration rendering

$$v(r) = \frac{r_e}{r}\frac{q_0}{\Phi(r)} \qquad \text{for } r_x \leq r \leq r_e \ . \tag{S10}$$

Fig. 2 in the main manuscript shows the flow velocity used in the simulation. The volume fraction in the symplast $\Phi(r)$ considers the periodic space restriction produced by vacoules and plasmodesmata. An estimate based on Rutschow et al. [2011] delivers a volume fraction in plasmodesmata of the order of 0.15, while the vacoule was assumed to make up 80% of the cell volume ($\Phi_3 = 0.2$). Table S.2 presents the geometry parameters on which the calculations were based. This parameter set represents a typical *A. thaliana* root.

## S.4 Zinc transport

Having determined the flow velocities in the compartments considered, we are able to move on to the task of finding a model for zinc. Its transport can be modelled by a diffusion advection problem, that states the conservation of zinc

$$\partial_t(\Psi\, \mathcal{Z}_i) - \mathrm{div}\,(\Psi\, D\, \mathrm{grad}\, \mathcal{Z}_i) = 0 \qquad \text{in } \Omega_i \times (0, \infty)\,,\ i = 1,\, 2\,, \qquad \text{(S11a)}$$

$$\partial_t(\Phi\, \mathcal{Z}_3) + \mathrm{div}\,(\Phi\, \mathbf{v}\, \mathcal{Z}_3 - \Phi\, D\, \mathrm{grad}\, \mathcal{Z}_3) = 0 \qquad \text{in } \Omega_3 \times (0, \infty)\,, \qquad \text{(S11b)}$$

where $\Psi$ is the volume fraction of the apoplasts $\Omega_1$ and $\Omega_2$, $\mathcal{Z}_i$ is the zinc concentration and $D$ a diffusion coefficient. Transport in the apoplast is assumed to take place only by means of diffusion, as the water flow velocity was assumed to be zero.

### S.4.1 Reduction to 1-D

Define $Z_i$ as the average of $\mathcal{Z}_i$ over $\Gamma_i(r)$:

$$Z_i(r, t) := \frac{1}{\mu_i(r)} \int_{\Gamma_i(r)} \mathcal{Z}_i(x, t)\, \mathrm{d}\gamma\,, \qquad \text{for } (r, t) \in [r_x, r_e] \times [0, \infty)\,.$$

We proceed to average Eqs. (S11a) and (S11b). We will average exemplarily Eq. (S11b). The result will apply also to Eq. (S11a) by exchanging $\Phi$ with $\Psi$ and setting $v = 0$. Treatment of the time derivative term is straightforward

$$\frac{1}{\mu_3(r)} \int_{\Gamma_3(r)} \partial_t\,(\Phi\, \mathcal{Z}_3)\, \mathrm{d}\gamma = \partial_t\,(\Phi\, Z_3)$$

The contribution of the advection term in Eq. (S11b) is

$$\frac{1}{\mu_3(r)} \int_{\Gamma_3(r)} \mathrm{div}(\Phi\, \mathbf{v}\, \mathcal{Z}_3)\, \mathrm{d}\gamma = \frac{1}{r}\, \partial_r\left(r\, \Phi\, \frac{1}{\mu_3(r)} \int_{\Gamma_3(r)} v_r\, \mathcal{Z}_3\, \mathrm{d}\gamma \right) + \frac{1}{\mu_3(r)} \int_{\partial \Gamma_3(r)} \Phi\, v_n\, \mathcal{Z}_3\, \mathrm{d}s\,,$$

where $v_n$ is the normal velocity on $\partial \Gamma_3(r)$. For the sake of generality the integral over $\partial \Gamma_3(r)$ is kept, although it is zero by the assumption of no water exchange between the apoplast and symplast ($v_n = 0$). By the mean value theorem, there exists a $\tilde{v}(r)$ so that

$$\int_{\Gamma_3(r)} v_r\, \mathcal{Z}_3\, \mathrm{d}\gamma = \tilde{v}(r) \int_{\Gamma_3(r)} \mathcal{Z}_3\, \mathrm{d}\gamma\,.$$

This renders the approximation

$$\frac{1}{\mu_3(r)} \int_{\Gamma_3(r)} \text{div}(\Phi\,\mathbf{v}\,\mathcal{Z}_3)\,d\gamma \approx \frac{1}{r}\,\partial_r\,(r\,\Phi\,v\,Z_3) + \frac{1}{\mu_3(r)} \int_{\partial\Gamma_3(r)} \Phi\,v_n\,\mathcal{Z}_3\,ds\;,$$

where $\tilde{v} \approx v$ was assumed. In the last step we had to make a somewhat crude approximation. However, without explicitely solving the whole 3-D problem, this is the best that can be done. The contribution of diffusion is obtained by expressing the divergence and gradient operators in cylindrical coordinates. The terms containing an $r$-derivative are

$$-\frac{1}{\mu_3(r)} \int_{\Gamma_3(r)} \frac{1}{r}\,\partial_r(\Phi\,D\,r\,\partial_r\mathcal{Z}_3)\,d\gamma = -\frac{1}{r}\,\partial_r\,(\Phi\,D\,r\,\partial_r Z_3)\;.$$

The term containing $\varphi$- and $z$-derivatives are transformed into an integral over the boundary of $\Gamma_3(r)$

$$-\frac{1}{\mu_3(r)} \int_{\Gamma_3(r)} \left(\frac{1}{r}\,\partial_\varphi\left(\frac{\Phi\,D}{r}\,\partial_\varphi\mathcal{Z}_3\right) + \partial_z(\Phi\,D\,\partial_z\mathcal{Z}_3)\right) d\gamma = -\frac{1}{\mu_3(r)} \int_{\partial\Gamma_3(r)} \Phi\,D\,\partial_n\mathcal{Z}_3\,ds\;,$$

where $\partial_n\mathcal{Z}_3$ is the normal derivative of $\mathcal{Z}_3$ on $\partial\Gamma_3(r)$. Summation of this boundary integral with the one obtained from the advection term delivers the average flux through the boundary

$$\frac{1}{\mu_3(r)} \int_{\partial\Gamma_3(r)} \Phi\,(v_n\,\mathcal{Z}_3 - D\,\partial_n\mathcal{Z}_3)\,ds = -Q_3\;,$$

$$Q_i := \sum_{j=1}^{2} \sigma_{ij}(r)\,J_j(r) \tag{S12}$$

where the flux densities $J_j(r)$ represent the transport through ZIP and HMA4 (developed in next section), $Q_3$ denominates the source of zinc, and $\sigma_{ij}$ is

$$(\sigma_{ij})(r) := \begin{pmatrix} -2/z_{0,1} & 0 \\ 0 & -2/r\varphi_{0,2} \\ 2/z_{0,3} & 2/r\varphi_{0,3} \end{pmatrix}\;. \tag{S13}$$

The sign convention of $J_i$ is as follows: positive for a flux from the apoplast into the symplast and negative vice versa. $\sigma_{ij}$ is related to the ratio of the length of the pieces composing the boundary $\partial\Gamma_3(r)$ to the area of $\Gamma_3(r)$ and weights, thus, the flux densities $J_j$ to account correctly the change in average concentration.

Putting everything together, the following reduced model for the symplast $\Omega_3$ is obtained

$$\partial_t(\Phi\,Z_3) + \frac{1}{r}\,\partial_r\,(r\,\Phi\,v\,Z_3 - \Phi\,D\,r\,\partial_r Z_3) = Q_3 \qquad \text{in } (r_x, r_e) \times (0, \infty)\;. \tag{S14}$$

It is practical for the implementation of a numerical method to rewrite the equation into a conservative form by introducing $r Z_i$ as a variable. Applying the results for $\Omega_3$ to $\Omega_1$ and $\Omega_2$, the following system is obtained

$$\begin{aligned}
\partial_t(\Psi\, r Z_1) + \partial_r\big((D/r)\, \Psi\, r Z_1 - \Psi\, D\, \partial_r(r Z_1)\big) &= r\, Q_1\,, \\
\partial_t(\Psi\, r Z_2) + \partial_r\big((D/r)\, \Psi\, r Z_2 - \Psi\, D\, \partial_r(r Z_2)\big) &= r\, Q_2\,, \quad \text{in } (r_x, r_e) \times (0, \infty)\,. \quad \text{(S15)} \\
\partial_t(\Phi\, r Z_3) + \partial_r\big((v + D/r)\, \Phi\, r Z_3 - \Phi\, D\, \partial_r(r Z_3)\big) &= r\, Q_3\,,
\end{aligned}$$

Initial values and boundary conditions are developed in Section S.4.3.

### S.4.2 Flux densities $J_j$

The zinc sources $Q_i$ on the right hand side of Eq. (S15) depend on the flux densities $J_j$, which still need to be specified. The zinc flux through transporters can be modelled by a saturable pointwise reaction mechanism

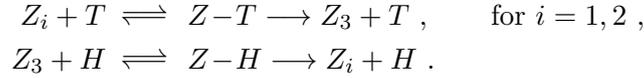

$$\begin{aligned}
Z_i + T &\rightleftharpoons Z\text{-}T \longrightarrow Z_3 + T\,, \qquad \text{for } i = 1, 2\,, \\
Z_3 + H &\rightleftharpoons Z\text{-}H \longrightarrow Z_i + H\,.
\end{aligned}$$

We will assume that the above reactions follow Michaelis-Menten kinetics and introduce as in Claus and Chavarría-Krauser [2012] a saturation function

$$f(Z, K) = \frac{Z}{Z + K}\,, \tag{S16}$$

where $K$ is the corresponding Michaelis-Menten constant. We introduce versions of $\mathcal{T}(x,t)$ and $\mathcal{H}(x)$ which depend on the radius by averaging over $\partial\Gamma_3(r)$

$$T(r,t) := \frac{1}{\mu(\partial\Gamma_3(r))} \int_{\partial\Gamma_3(r)} \mathcal{T}(x,t)\, \mathrm{d}s\,, \quad \text{for } (r,t) \in [r_x, r_e] \times [0, \infty)\,, \tag{S17a}$$

$$H(r) := \frac{1}{\mu(\partial\Gamma_3(r))} \int_{\partial\Gamma_3(r)} \mathcal{H}(x)\, \mathrm{d}s\,, \quad \text{for } r \in [r_x, r_e]\,. \tag{S17b}$$

Note that, $T(r,t)$ is equal to $T_{en}(t)$, $T_{co}(t)$ and $T_{ep}(t)$, respectively, for an $r$ inside one of these cells, zero elsewhere. $H(r)$ is equal to $H_{pc}$ for an $r$ inside the pericycle and zero elsewhere. The reaction probability depends on $\Psi T(r,t)$ instead of only $T(r,t)$, because $\Omega_1$ and $\Omega_2$ are porous media and only the reduced amount $\Psi T(r,t)$ has actually contact to $Z_1$ and $Z_2$. No correction is needed for $H(r)$, as the cytoplasm can be assumed to have direct contact with the membrane, so that the complete $H(r)$ can react with $Z_3$.

In total, the flux densities $J_j$ are modelled as

$$J_j(r,t) = \Phi\, T_0\, T(r,t)\, f\big(Z_j(r,t), K^t\big) - H_0\, H(r)\, f\big(Z_3(r,t), K^h\big)\,, \qquad j = 1, 2\,, \tag{S18}$$

where $T_0$ and $H_0$ are constants that characterize the true amount of transporters (non-dimensionalized regulation).

### S.4.3 Initial values and boundary conditions

Equation (S15) needs initial values and suitable boundary contions to obtain a well posed (solvable) problem. The apoplast is assumed to have access to a perfectly stirred medium of concentration $Z^e$. A concentration $Z^x(t)$ is prescribed at the xylem. This concentration depends on the flux of zinc through HMA4 and a model will be developed in the next section. The impermeability of the Casparian strip is considered by setting a no-flux condition. In total, we prescribe for the apoplasts $\Omega_1$ and $\Omega_2$

$$
\begin{aligned}
Z_i\Big|_{r=r_x} &= Z^x(t) \,, \\
(D/r)\,\Psi\,rZ_i - \Psi\,D\,\partial_r(rZ_i)\Big|_{r\nearrow r_c} &= 0 \,, \\
(D/r)\,\Psi\,rZ_i - \Psi\,D\,\partial_r(rZ_i)\Big|_{r\searrow r_c} &= 0 \,, \\
Z_i\Big|_{r=r_e} &= Z^e \,,
\end{aligned}
\qquad \text{for } t \in [0,\infty),\ i=1,2\,. \qquad (S19)
$$

The zinc flux is prescribed at the boundary of the symplast $\Omega_3$

$$
(v + D/r)\,\Phi\,rZ_3 - \Phi\,D\,\partial_r(rZ_3)\Big|_{r=r_x} = r\,H_0\,H_{pc}\,f\bigl(Z_3(r_x), K^h\bigr) \,, \qquad (S20a)
$$

$$
(v + D/r)\,\Phi\,rZ_3 - \Phi\,D\,\partial_r(rZ_3)\Big|_{r=r_e} = -r\,\Phi\,T_0\,T_{ep}\,f\bigl(Z^e, K^t\bigr) \,, \qquad (S20b)
$$

$$
\text{for } t \in [0, \infty) \,.
$$

Published experimental results that capture the dynamics of regulation, focus on changes from one steady state at a given external concentration to another steady state for a different concentration (e.g. zinc resupply). Therefore, the initial conditions used here are solutions of the stationary version of Eq. (S15)

$$
Z_i(r,t)\Big|_{t=0} = \overline{Z}_i(r) \qquad \text{for } r \in [r_x, r_e]\,,\ i=1,2,3, \qquad (S21)
$$

where the $\overline{Z}_i$ fulfill the following equations

$$
\partial_r\bigl((D/r)\,\Psi\,r\overline{Z}_i - \Psi\,D\,\partial_r(r\overline{Z}_i)\bigr) = r\,Q_i \,, \quad \text{in } (r_x, r_e) \times (0,\infty)\,,\ i=1,2\,, \qquad (S22)
$$

$$
\partial_r\bigl((v + D/r)\,\Phi\,r\overline{Z}_3 - \Phi\,D\,\partial_r(r\overline{Z}_3)\bigr) = r\,Q_3 \quad \text{in } (r_x, r_e) \times (0,\infty) \,, \qquad (S23)
$$

with boundary conditions Eqs. (S19), (S20a) and (S20b). In general, the steady state for given $Z^e$ and $q_0$ will be determined, used as a initial condition, one of these parameters changed, and the dynamics of the transition captured.

### S.4.4 Xylem

A calculation of the apoplastic zinc concentration in the region enclosed by the Casparian strip ($r_x \leq r \leq r_c$) needs the concentration of zinc in the xylem. For simplicity, we will pose a model for the central cylinder $0 \leq r < r_x$ (i.e. stele without the pericycle) and account the true size of the xylem by a constant volume fraction $\Psi_x$. The domain describing this tissue will be denoted as $\Omega_x$, where the $x$ stands for *xylem*.

Eqs. (S5) and (S11b) apply also to this tissue

$$\mathrm{div}(\Psi_x \mathbf{v}_x) = 0 \qquad \text{in } \Omega_x \, , \tag{S24}$$

$$\partial_t(\Psi_x \, \mathcal{Z}_x) + \mathrm{div}\left(\Psi_x \, \mathbf{v}_x \, \mathcal{Z}_x - \Psi_x \, D \, \mathrm{grad} \, \mathcal{Z}_x\right) = 0 \qquad \text{in } \Omega_x \times (0, \infty) \, , \tag{S25}$$

where $\mathbf{v}_x$ is the flow velocity and $\mathcal{Z}_x$ is the zinc concentration in the central cylinder $\Omega_x$. An average can be obtained as in Secs. S.3 and S.4. The main difference is that the surface over which the average is created is here

$$\Gamma_x(z) := \{ \mathbf{x} \in \Omega_x \mid x_3 = z \} \, .$$

Conservation of water delivers

$$\partial_z(\Psi_x \, v_x) = -\frac{2}{r_x} \Phi(r_x) \, v(r_x) \, , \tag{S26}$$

where $\Phi(r_x) v(r_x)$ is the flow velocity of the water being delivered from the symplast, and the average velocity in the xylem is defined as

$$v_x(z) := \frac{1}{\mu(\Gamma_x(z))} \int_{\Gamma_x(z)} v_{z,x}(r, \varphi, z) \, \mathrm{d}\gamma \, , \qquad \text{for } z \in [0, L] \, ,$$

where $L$ is the length of the root portion considered. Expression of $\Phi(r_x) \, v(r_x)$ by Eq. (S10), integration of Eq. (S26), and assumption of $v_x(0) = 0$ delivers

$$v_x(z) = -\frac{2}{r_x} \frac{r_e}{r_x} \frac{q_0}{\Psi_x} z \qquad \text{for } z \in [0, L] \, , \tag{S27}$$

which is a linear function of $z$. Rememeber that $q_0 < 0$ so that $v_x(z) \geq 0$ for $z \geq 0$. Eq. (S27) is based on the assumption that $q_0$ is constant, which will not be true in reality. The pressure gradient between the xylem and the medium will fall with $z$ and, hence, $v_x(z)$ cannot grow linearly indefinitelly and will stagnate at a constant value. However, $v_x(z)$ will behave similar to Eq. (S27) in a region near $z = 0$. We focus on this region and assume validity of Eq. (S27).

An average for the equation describing the conservation of zinc is obtained readily

$$\partial_t(\Psi_x Z_x) + \partial_z(\Psi_x \, v_x \, Z_x - \Psi_x \, D \, \partial_z Z_x) = \frac{2}{r_x} H_0 \, H_{pc} \, f\bigl(Z_3(r_x, t), K^h\bigr) \, , \tag{S28}$$

where the boundary condition Eq. (S20a) divided by $r$ was used and the average zinc concentration is defined as

$$Z_x(z, t) := \frac{1}{\mu(\Gamma_x(z))} \int_{\Gamma_x(z)} \mathcal{Z}_x(x, t) \, \mathrm{d}\gamma \qquad \text{for } (z, t) \in [0, L] \times [0, \infty) \, .$$

A preliminary simulation of this equation with no-flux and open vessel conditions at $z = 0$ and $z = L$, respectivelly, shows that $Z_x$ is almost constant in space. Hence, we set $Z_x(z,t) \approx Z^x(t)$, use that $\Psi_x$ is constant and express $v_x$ by Eq. (S27) to obtain

$$\frac{dZ^x}{dt} = \frac{1}{\Psi_x} \frac{2}{r_x} \left( \frac{r_e}{r_x} q_0 Z^x + H_0 H_{pc} f\big(Z_3(r_x, t), K^h\big) \right) \qquad \text{for } t \in (0, \infty), \qquad (S29)$$
$$Z^x\big|_{t=0} = Z_0^x.$$

Note again that $q_0 < 0$, so that this equation has a non-trivial positive steady state solution

$$\overline{Z}^x = -\frac{r_x}{r_e} \frac{H_0 H_{pc}}{q_0} f\big(\overline{Z}_3(r_x), K^h\big). \qquad (S30)$$

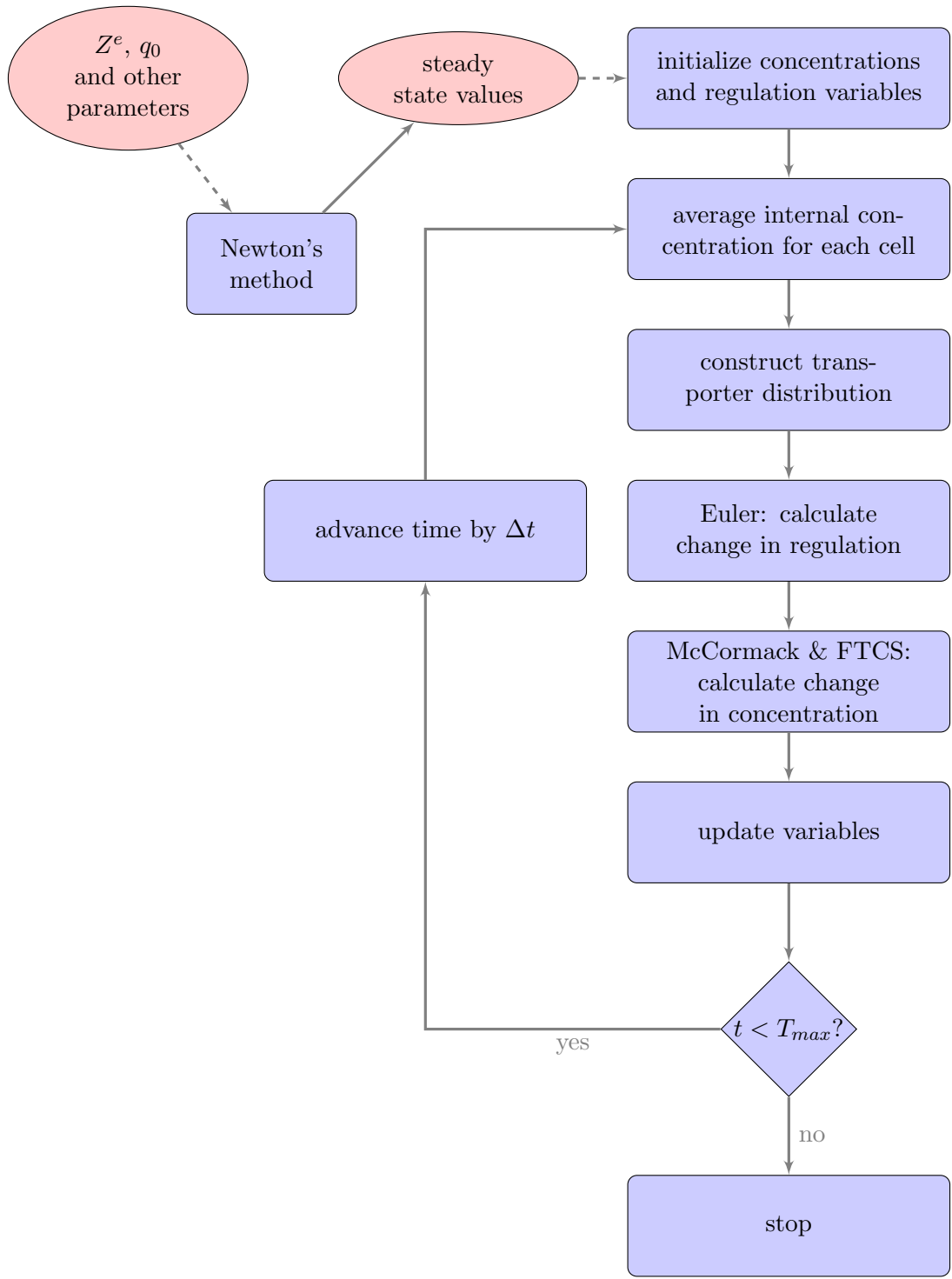

Figure S.2: Diagram of the steps involved in coupling the numerical schemes.

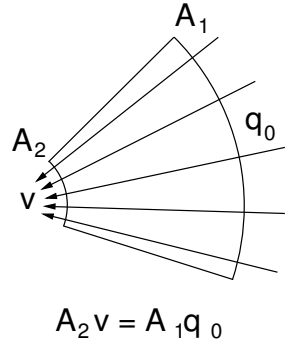

Figure S.3: Schematic diagram of effect of cylindrical geometry on water flow velocity. The flow cross section becomes smaller for smaller radii resulting in acceleration of the flow: $v = \frac{A_1}{A_2} q_0 > q_0$ for $A_1 > A_2$.

Table S.1: Parameters used in the regulation model. Values taken from Claus and Chavarría-Krauser [2012]

| Parameter | $K^{t*}$ | $K$ | $\Gamma$ | $\Gamma'$ | $\Gamma_I$ | $\gamma_G$ | $\gamma_M$ | $\gamma_T$ | $\gamma_A$ | $\gamma_I$ |
|---|---|---|---|---|---|---|---|---|---|---|
| Value | 13 μM | 20 | 38 | 167.2 | 1000 | 4 | 4 | 1 | 1 | 1 |

* Value for ZIP1, Grotz et al. [1998]

Table S.2: Geometry parameters used in the simulation. These correspond to a typical *Arabidopsis thaliana* root.

| Parameter | $r_x$ | $r_c$ | $r_e$ | $z_{0,1}$ | $z_{0,3}$ | $\phi_{0,3}$ |
|---|---|---|---|---|---|---|
| Value | 6 μm | 12.5 μm | 40 μm | 0.5 μm | 135 μm | $\pi/10$ |

Table S.3: Further parameters used in the simulation.

| Parameter | Value | Description |
|---|---|---|
| $K^h$ | 1 μM | Michaelis-Menten constant for HMA4 |
| $\zeta_0$ | 166.67 μM | Scaling factor that dimensionalizes the internal zinc concentration |